\documentclass[twocolumn,showpacs,aps,prl,preprintnumbers,superscriptaddress]{revtex4}

\usepackage{graphicx}
\usepackage{dcolumn}
\usepackage{amsmath}
\usepackage{epsfig}

\long\def\inst#1{\par\nobreak\kern 4pt\nobreak
    {\itshape #1}\par\vskip 10pt plus 3pt minus 3pt}
\RequirePackage{xspace}
\usepackage{relsize}

\def\fL   {\ensuremath{ f_L }}

\def\qqbar {\ensuremath{q\overline q}\xspace}
\def\babar{\mbox{\slshape B\kern-0.1em{\smaller A}\kern-0.1em
    B\kern-0.1em{\smaller A\kern-0.2em R}}}
\def\Dbar    {\kern 0.18em\overline{\kern -0.18em D}{}\xspace}
\def\Bbar    {\kern 0.18em\overline{\kern -0.18em B}{}\xspace}

\def\BB      {\ensuremath{B\Bbar}\xspace} 
\def\Bz      {\ensuremath{B^0}\xspace}
\def\Bzb     {\ensuremath{\Bbar^0}\xspace}
\def\BzBzb   {\ensuremath{\Bz {\kern -0.16em \Bzb}}\xspace}
\def\Bu      {\ensuremath{B^+}\xspace}
\def\Bub     {\ensuremath{B^-}\xspace}

\def\Bm      {\ensuremath{\Bub}\xspace}
\def\Bpm     {\ensuremath{B^\pm}\xspace}

\def\BpBm    {\ensuremath{\Bu {\kern -0.16em \Bub}}\xspace}

\newcommand{\optbar}[1]{\shortstack{{\tiny (\rule[.4ex]{1em}{.1mm})}
  \\ [-.7ex] $#1$}}
\def\BorBbar    {\kern 0.18em\optbar{\kern -0.18em B}{}\xspace}
\def\DorDbar    {\kern 0.18em\optbar{\kern -0.18em D}{}\xspace}
\def\KorKbar    {\kern 0.18em\optbar{\kern -0.18em K}{}\xspace}

\def\pep2{PEP-II}
\mathchardef\Upsilon="7107
\def\Y#1S{\ensuremath{\Upsilon{(#1S)}}\xspace}
\def\FourS {\Y4S}
\def\qqbar {\ensuremath{q\overline q}\xspace}
\def\pip   {\ensuremath{\pi^+}\xspace}
\def\pim   {\ensuremath{\pi^-}\xspace}
\def\pipi  {\ensuremath{\pi^+\pi^-}\xspace}
\def\pipm  {\ensuremath{\pi^\pm}\xspace}

\def\BF{$B$ Factory}
\def\invfb   {\ensuremath{\mbox{\,fb}^{-1}}\xspace}

\def\B       {\ensuremath{B}\xspace}
\def\mes        {\mbox{$m_{\rm ES}$}\xspace}
\def\DeltaE     {\mbox{$\Delta E$}\xspace}
\def\Dm      {\ensuremath{D^-}\xspace}
\def\Dz      {\ensuremath{D^0}\xspace}
\def\Kp    {\ensuremath{K^+}\xspace}
\def\Km    {\ensuremath{K^-}\xspace}

\newcommand{\jprlBase}       {Phys.\ Rev.\ Lett.\xspace}
\newcommand{\jprl}      [1]  {\jprlBase\ {\bf #1}}
\newcommand{\jplBase}        {Phys.\ Lett.\xspace}
\newcommand{\jpl}       [1]  {\jplBase\ {\bf #1}}
\newcommand{\plb}       [1]  {\jplBase\ B~{\bf #1}}
\newcommand{\jprBase}        {Phys.\ Rev.\xspace}
\newcommand{\jprd}      [1]  {\jprBase\ D~{\bf #1}}
\newcommand{\npBase}         {Nucl.\ Phys.\xspace}
\newcommand{\npb}       [1]  {\npBase\ B~{\bf #1}}
\newcommand{\nimBaseA}       {Nucl.\ Instrum.\ Methods Phys.\ Res., Sect.\ A\xspace}
\newcommand{\nima}      [1]  {\nimBaseA~{\bf #1}}

\def\pipi   {\ensuremath{\pi\pi}}

\def\mkpi   {\ensuremath{m_{K\pi}}}

\def\VV     {\ensuremath{VV}\xspace}

\def\calF      {\ensuremath{{\cal F}}\xspace}

\def\Kbar  {\kern 0.2em\overline{\kern -0.2em K}{}\xspace}

\def\Kz    {\ensuremath{K^0}\xspace}
\def\Kzb   {\ensuremath{\Kbar^0}\xspace}
\def\Kstarz  {\ensuremath{K^{*0}}\xspace}
\def\Kstarzb {\ensuremath{\Kbar^{*0}}\xspace}
\def\KorKbar   {\kern 0.18em\optbar{\kern -0.18em K}{}\xspace}

\newcommand{\stat}{\ensuremath{\mathrm{(stat)}}\xspace}
\newcommand{\syst}{\ensuremath{\mathrm{(syst)}}\xspace}

\def\KstarzKstarzb    {\ensuremath{\Kstarz \Kstarzb}}
\def\btoKstarzKstarzb {\ensuremath{\Bz \rightarrow \Kstarz \Kstarzb}}

\def\KstarzKstarz     {\ensuremath{\Kstarz\Kstarz}}
\def\KstarzbKstarzb     {\ensuremath{\Kstarzb\Kstarzb}}

\def\btoKstarzKstarz {\ensuremath{\Bz \rightarrow \Kstarz \Kstarz}}

\def\KstarzII  {\ensuremath{K^{*0}(1430)}\xspace}
\def\KstarzbII {\ensuremath{\Kbar^{*0}(1430)}\xspace}

\newcommand{\onreslumi} {\mbox{348\invfb}}
\newcommand{\offreslumi} {\mbox{36.6\invfb}}
\newcommand{\nbb}       {\mbox{$383.2\pm4.2$}}

\newcommand{\gev}{\ensuremath{\mathrm{\,Ge\kern -0.1em V}}\xspace}
\newcommand{\mev}{\ensuremath{\mathrm{\,Me\kern -0.1em V}}\xspace}
\newcommand{\kev}{\ensuremath{\mathrm{\,ke\kern -0.1em V}}\xspace}
\newcommand{\ev}{\ensuremath{\mathrm{\,e\kern -0.1em V}}\xspace}
\newcommand{\gevc}{\ensuremath{{\mathrm{\,Ge\kern -0.1em V\!/}c}}\xspace}
\newcommand{\mevc}{\ensuremath{{\mathrm{\,Me\kern -0.1em V\!/}c}}\xspace}
\newcommand{\gevcc}{\ensuremath{{\mathrm{\,Ge\kern -0.1em V\!/}c^2}}\xspace}
\newcommand{\mevcc}{\ensuremath{{\mathrm{\,Me\kern -0.1em V\!/}c^2}}\xspace}
\def\etal{{\em et al.}}
\def\epem       {\ensuremath{e^+e^-}\xspace}

\def\cossq    {\ensuremath{\cos^2\theta}}
\def\sinsq    {\ensuremath{\sin^2\theta}}

\def\Bmeson  {\B\ meson}
\def\Bmesons {\B\ mesons}

\def\Bback   {\BB\ background}
\def\Bbacks  {\BB\ backgrounds}

\newcommand{\kzbnsig}     {\mbox{$33.5^{+9.1}_{-8.1}$}}
\newcommand{\kzbbfcor}    {\mbox{$1.28^{+0.35}_{-0.30} \pm 0.11$}}
\newcommand{\kzbbfcorsys} {\mbox{$1.28^{+0.35}_{-0.30} \stat \pm 0.11 \syst$}}

\newcommand{\kzbfl}    {\mbox{$0.80^{+0.10}_{-0.12} \pm 0.06$}}
\newcommand{\kzbflsys} {\mbox{$0.80^{+0.10}_{-0.12}\ \stat \pm 0.06\ \syst$}}

\newcommand{\kzbsig}   {\mbox{$6$}} 
\newcommand{\kznsig}   {\mbox{$2.7 \pm 3.3$}}
\newcommand{\kzbfcor}  {\mbox{$0.11^{+0.16}_{-0.11} \pm 0.04$}}

\newcommand{\kzfl}     {\mbox{$1.0 \pm 1.0$}}
\newcommand{\kzupcor}  {\mbox{$0.41$}}
\newcommand{\kzsig}    {\mbox{$0.9$}} 

\begin{document}

\preprint{\babar-PUB-07/050}
\preprint{SLAC-PUB-12724}

\title{
\large \bfseries \boldmath
Observation of \btoKstarzKstarzb\ and 
search for \btoKstarzKstarz
}

\date{\today}
%
\author{B.~Aubert}
\author{M.~Bona}
\author{D.~Boutigny}
\author{Y.~Karyotakis}
\author{J.~P.~Lees}
\author{V.~Poireau}
\author{X.~Prudent}
\author{V.~Tisserand}
\author{A.~Zghiche}
\affiliation{Laboratoire de Physique des Particules, IN2P3/CNRS et Universit\'e de Savoie, F-74941 Annecy-Le-Vieux, France }
\author{J.~Garra~Tico}
\author{E.~Grauges}
\affiliation{Universitat de Barcelona, Facultat de Fisica, Departament ECM, E-08028 Barcelona, Spain }
% ...
\collaboration{The \babar\ Collaboration}
\noaffiliation

\begin{abstract}
We report the observation of the $b\rightarrow d$ penguin-dominated
decay \btoKstarzKstarzb\ with a sample of \nbb\ million \BB\ pairs
collected with the \babar\ detector at the PEP-II asymmetric-energy
\epem\ collider at the Stanford Linear Accelerator Center. 
The measured branching fraction is ${\cal B}
(\btoKstarzKstarzb) = [\kzbbfcor] \times 10^{-6}$ and the fraction of
longitudinal polarization is $\fL(\btoKstarzKstarzb)$ = \kzbfl. The
first error quoted is statistical and the second systematic. We
also obtain an upper limit at the 90\% confidence level on the branching
fraction for ${\cal B} (\btoKstarzKstarz) < \kzupcor\times 10^{-6}$.
\end{abstract}


\maketitle


The study of the branching fractions and angular distributions of
\Bmeson\ decays to hadronic final states without a charm quark probes
the dynamics of both weak and strong interactions, and plays an
important role in understanding $CP$ violation. Decays proceeding via
electroweak and gluonic $b\rightarrow d$ penguin diagrams have only
recently been measured in the decays
$\B\rightarrow\rho\gamma$~\cite{bib:rhogamma} and
$\Bz\rightarrow\Kz\Kzb$~\cite{bib:kzkzbar}.  On the
other hand, the charmless decay \btoKstarzKstarzb\ proceeds through
both electroweak and gluonic $b\rightarrow d$ penguin loops to two
vector particles (\VV). The Standard Model (SM) suppressed decay
\btoKstarzKstarz\ could appear via an intermediate heavy boson.

Theoretical models in the framework of QCD factorization predict the
angular distribution of the \VV\ decays of the \Bmeson, as measured by
the longitudinal polarization fraction \fL, to be $\sim 0.9$ for
both tree- and penguin-dominated
decays~\cite{bib:prediction}. However, recent measurements of the pure
penguin \VV\ decay $B\rightarrow \phi K^*$ indicate \fL $\sim
0.5$~\cite{bib:phiKst}.  Several attempts to understand this
unexpected value of \fL\ within or beyond the Standard Model have
been made~\cite{bib:theory1}.  Further information about decays
related by $SU(3)$ symmetry may provide insight into this
polarization puzzle and test factorization models.  A time-dependent
angular analysis of \btoKstarzKstarzb\ can distinguish between penguin
annihilation and rescattering as mechanisms for the value of \fL\ 
observed in $B\rightarrow \phi K^*$~\cite{bib:datta}. The
\btoKstarzKstarzb\ mode can also be used within the SM framework to
help constrain the angles $\alpha$ and $\gamma$ of the Unitarity
Triangle~\cite{bib:theory2}.

Theoretical calculations for \btoKstarzKstarzb\ branching fractions
 cover the range $(0.16\mbox{-}0.96)\times 10^{-6}$~\cite{bib:all}.
 Recently, Beneke, Rohrer, and Yang~\cite{bib:Beneke06} predicted
 $(0.6^{+0.1 +0.3}_{-0.1 -0.2}) \times 10^{-6}$ and $\fL\ = 0.69 \pm
 0.01 ^{+0.16}_{-0.20}$. Experimentally, upper limits on the branching
 fractions at the 90\% confidence level (C.L.)  of $22\times 10^{-6}$
 and $37\times 10^{-6}$ exist for \btoKstarzKstarzb\ and
 \btoKstarzKstarz, respectively~\cite{bib:prevcleo}.


We report measurements of the branching fraction and the fraction of
 longitudinal polarization for the decay mode \btoKstarzKstarzb, with
 explicit consideration of non-resonant backgrounds and interference
 from \Kstarz\KstarzbII. We place an upper limit on the branching
 fraction of \btoKstarzKstarz, where we use the notation
 \KstarzKstarz\ to also represent \KstarzbKstarzb.
 Charge-conjugate modes are implied throughout and we assume equal
 production rates of \BpBm and \BzBzb.

This analysis is based on a data sample of \nbb\ million \BB\ pairs,
corresponding to an integrated luminosity of \onreslumi, collected
with the \babar\ detector at the PEP-II
asymmetric-energy \epem\ collider operated at the Stanford Linear
Accelerator Center. The \epem\ center-of-mass
(c.m.) energy is $\sqrt{s} = 10.58$\gev, corresponding to the
\FourS\-resonance mass (on-resonance data). In addition, \offreslumi\
of data collected 40~\mev\ below the \FourS\-resonance (off-resonance
data) are used for background studies.


The \babar\ detector is described in detail in Ref.~\cite{bib:babar}.
Charged particles are reconstructed as tracks with a 5-layer silicon
vertex detector and a 40-layer drift chamber inside a $1.5$-T
solenoidal magnet. An electromagnetic calorimeter is used to
identify electrons and photons.  A ring-imaging Cherenkov detector
 is used to identify charged hadrons and provides additional
electron identification information. Muons are identified by an
instrumented magnetic-flux return.


The \btoKstarzKstarzb\ and \btoKstarzKstarz\ candidates are 
reconstructed through the decays $\Kstarz\to K^+\pi^-$ and $\Kstarzb\to
K^-\pi^+$. The differential decay rate, after
integrating over the angle between the decay planes of the vector
mesons, for which the acceptance is uniform, is
\begin{equation}
\frac{1}{\Gamma}\frac{d^2\Gamma}{d\cos\theta_{1}d\cos\theta_{2}} 
\propto \frac{1-f_L}{4}\sin^2\theta_{1}\sin^2\theta_{2} + 
   f_L \cos^2\theta_{1}\cos^2\theta_{2} ,
\label{eq:helicity}
\end{equation}

\noindent where $\theta_{1}$ and $\theta_{2}$ are the helicity angles
of the \Kstarz\ or \Kstarzb.  The helicity angle of the $\Kstarz\
(\Kstarzb)$ is defined as the angle between the $K^+(K^-)$ momentum
and the direction opposite to the \Bmeson\ in the \Kstarz(\Kstarzb)
rest frame~\cite{bib:polarization}.



The charged tracks from the \Kstarz\ decays are required to have at
least 12 hits in the drift chamber and a transverse momentum greater
than 0.1\gevc. The tracks are identified as either pions or kaons
 by measurement of the energy loss in the tracking
devices, the number of photons measured by the Cherenkov detector
and the corresponding Cherenkov angles. These measurements are combined
with calorimeter information to reject electrons, muons, and protons.
We require the invariant mass of the \Kstarz\  candidates to be $0.792
< \mkpi < 1.025$\gevcc. A \Bmeson\ candidate is formed from two
\Kstarz\ candidates, with the constraint that the two \Kstarz\ 
candidates originate from the interaction region.

\Bmeson\ candidates are characterized kinematically by the energy
 difference $\DeltaE = E^*_B - \sqrt{s}/2$ and the energy-substituted
 mass $m_{\rm ES} = \left [ (s/2+{\bf p}_i\cdot{\bf p}_B)^2/E_i^2-{\bf
 p}_B^2\right ] ^{1/2}$, where $(E_i,{\bf p}_i)$ and $(E_B,{\bf p}_B)$
 are the four-momenta of the \FourS and \Bmeson\ candidate,
 respectively, and the asterisk denotes the \FourS\ rest frame. The
 total event sample is taken from the region $-0.08 \le \DeltaE \le
 0.2$\gev and $5.25 \le \mes \le 5.29$\gevcc. Events outside the
 region $|\DeltaE|\le 0.07\gev$ and $5.27 \le \mes \le 5.29\gevcc$ are
 used to characterize the background.  The average number of signal
 \Bmeson\ candidates per selected data event is 1.03.  A single
 \Bmeson\ candidate per event is chosen as the one whose fitted decay
 vertex has the smallest $\chi^2$. MC simulations show that up to 4\%
 (1.6\%) of longitudinally (transversely) polarized signal events are
 misreconstructed, with one or more tracks originating from the other
 \Bmeson\ in the event.

To reject the dominant background consisting of light-quark \qqbar\
$(q = u,d,s,c)$ continuum events, we require $|\cos\theta_T|<0.8$,
where $\theta_T$ is the angle, in the c.m.\ frame, between the thrust
axes~\cite{bib:thrust} of the \Bmeson\ and that formed from the other
tracks and neutral clusters in the event. We create a Fisher
discriminant ${\cal F}$ to be used in the maximum-likelihood (ML) fit,
constructed from a linear combination of five variables: the polar
angles of the \Bmeson\ momentum vector and the \Bmeson\ thrust axis
with respect to the beam axis, the ratio of the second- and
zeroth-order momentum-weighted Legendre polynomial moments of the
energy flow around the \Bmeson\ thrust axis in the c.m.\
frame~\cite{bib:Legendre}, the flavor of the other \Bmeson\ as
reported by a multivariate tagging algorithm~\cite{bib:tagging}, and
the boost-corrected proper-time difference between the decays of the
two \Bmesons\ divided by its variance. The second \Bmeson\ is formed
by creating a vertex from the remaining tracks that are consistent
with originating from the interaction region.

We suppress background from decays to charmed states by removing
 candidates that have decays consistent with $\Dm\to
\Kp\pim\pim$ and an invariant mass in the range
 $1.845 < m_{\Kp\pim\pim} < 1.895$\gevcc. 
We reduce backgrounds from $\Bz\to \phi\Kstarz$
by assigning the kaon mass to the pion candidate and rejecting the
event if the combined invariant mass of the two charged tracks is
between $1.00$ and $1.04$\gevcc.  Finally, we
require the cosine of the helicity angle of both \Kstarz\  candidates to be
less than 0.98 to reduce continuum background and avoid the region
where the reconstruction efficiency falls off rapidly.


We use an extended unbinned ML fit to extract
the signal yield and polarization
simultaneously for each mode. The extended likelihood function is
\begin{equation}
{\mathcal L} = \frac{1}{N!}\exp{\left(-\sum_{j}n_{j}\right)}
\prod_{i=1}^N\left[\sum_{j}n_{j}{\mathcal
    P}_{j}(\vec{x}_i;\vec{\alpha}_j)\right]\!.
\end{equation}

\noindent We define the likelihood ${\cal L}_i$ for each event
candidate $i$ as the sum of $n_j {\cal P}_j(\vec x_i; \vec \alpha_j)$
over four hypotheses $j$ (signal, \qqbar\ background, \KstarzII\ and
\Bbacks\ as discussed below), where ${\cal P}_j(\vec x_i; \vec
\alpha_j)$ is the product of the probability density functions (PDFs)
for hypothesis $j$ evaluated for the $i$-th event's measured variables
$\vec x_i$, $n_j$ is the yield for hypothesis $j$, and $N$ is the
total number of events in the sample. The quantities $\vec \alpha_j$
represent parameters in the expected distributions of the measured
variables for each hypothesis $j$.  Each discriminating variable $\vec
x_i$ in the likelihood function is modeled with a PDF, where the
parameters $\vec \alpha_j$ are extracted from MC simulation,
off-resonance data, or (\mes, \DeltaE) sideband data.

The seven variables $\vec x_i$ used in the fit are \mes, \DeltaE,
\calF, and the invariant masses and cosines of the helicity angle of
the two \Kstarz\ candidates. Since the correlations among the fitted
input variables are found to be on average $\sim 1\%$ with a maximum
of 4\%, we take each ${\cal P}_j$ to be the product of the PDFs for
the separate variables. The effect of neglecting correlations is
evaluated by fitting ensembles of simulated experiments in which we
embed signal and background events randomly extracted from
fully-simulated MC samples.

The two invariant mass and helicity angle distributions for each
 \Kstarz\ meson are indistinguishable and so we use the same PDF
 parameters for both \Kstarz\ candidates. Peaking PDF distributions
 are described with an asymmetric Gaussian or a sum of two Gaussians.
 The transverse (longitudinal) helicity angle distributions are
 described with a \cossq\ (\sinsq) function corrected for changes in
 efficiency as a function of helicity angle.  The \Bbacks\ use an
 empirical non-parametric function for \DeltaE, the masses and
 helicity angles. The continuum background \mes\ shape is described by
 the function $x\sqrt{1-x^2}\exp[-\xi (1-x^2)]$ (with $x=\mes/E^*_B$
 and $\xi$ a free parameter)~\cite{bib:argus} and a first- or
 third-order polynomial is used for \DeltaE\ and the helicity angles,
 respectively. The continuum invariant mass distributions contain real
 \Kstarz\ candidates; we model the peaking mass component using the
 parameters extracted from the fit to the signal invariant mass
 distributions together with a second-order polynomial to represent
 the non-peaking component.

We use the decay $\Bz\rightarrow \Dm\pip (\Dm\to\Kstarz\pim)$ as a
calibration channel to account for small differences between MC
simulation and reconstructed data.  This decay has a similar topology
to the modes under study and is selected using the same criteria as
for \KstarzKstarzb\ but requiring the reconstructed \Kstarz\pipm\
invariant mass to be in the range $1.845 < m_{\Kstarz\pipm} <
1.895$\gevcc.  We predict $1860\pm186$ signal events and measure
$1614\pm47$.

We use MC-simulated events to study backgrounds from other \Bmeson\ 
decays. The major charmless \Bback\ to
\btoKstarzKstarzb\ is $\Bz\to \phi\Kstarz$, while charm \Bbacks\
are effectively suppressed by the requirement that the two pions (and
kaons) have opposite charge. For \btoKstarzKstarz, $\Bz\to
\phi\Kstarz$ remains the major charmless \Bback, but a number
of charm decays contaminate the signal, dominated by decays of the
type $\Bz\to \Dm\Kp$ and $\Bm\to \Dz\Km$. Given the uncertainty in the
polarization and branching fractions of these backgrounds, we allow
the \Bback\ yield to float in the fit.

A possible background is the decay $\Bz\to\Kstarz\KstarzbII$. We use
the LASS parameterization for the \KstarzbII\ lineshape, which
consists of the \KstarzbII\ resonance together with an effective-range
non-resonant component~\cite{bib:lass}. We apply the same selection
criteria used for \KstarzKstarzb\ but require one of the \Kstarz\
candidates to have an invariant mass in the range $1.025 < \mkpi <
1.53$\gevcc\ and perform an extended unbinned ML fit with the four
variables \mes, \DeltaE, \calF, and the \Kstarz\ mass. We fit the LASS
parameterization to the selected signal events in the \KstarzbII\ mass
range and extrapolate to the \Kstarz\ mass range. Interference effects
between the \Kstarz\ and the spin-0 final states (non-resonant and
\KstarzbII) integrate to zero as the acceptance of the detector and
analysis is uniform. Assuming no interference, we expect $6\pm5$
$\Bz\to\Kstarz\KstarzbII$ events in the fitted \btoKstarzKstarzb\
signal region. The uncertainty on the contribution is calculated from
the statistical error and the large uncertainty in the fitted LASS
parameters used to describe the \KstarzbII\ lineshape. We fix the
yield in the final fit and vary the yield by its error to assess the
systematic uncertainty.

The continuum background PDF parameters that are allowed to vary are
the \calF\ peak position, $\xi$ for \mes, the slope of \DeltaE, and
the polynomial coefficients and normalization describing the mass and
helicity angle distributions. We fit for ${\cal B}$ and \fL\ directly
and exploit the fact that ${\cal B}$ is less correlated with \fL\ than
is either the yield or efficiency taken separately.

\begin{figure}[t]
\centerline{
\setlength{\epsfxsize}{0.5\linewidth}\leavevmode\epsfbox{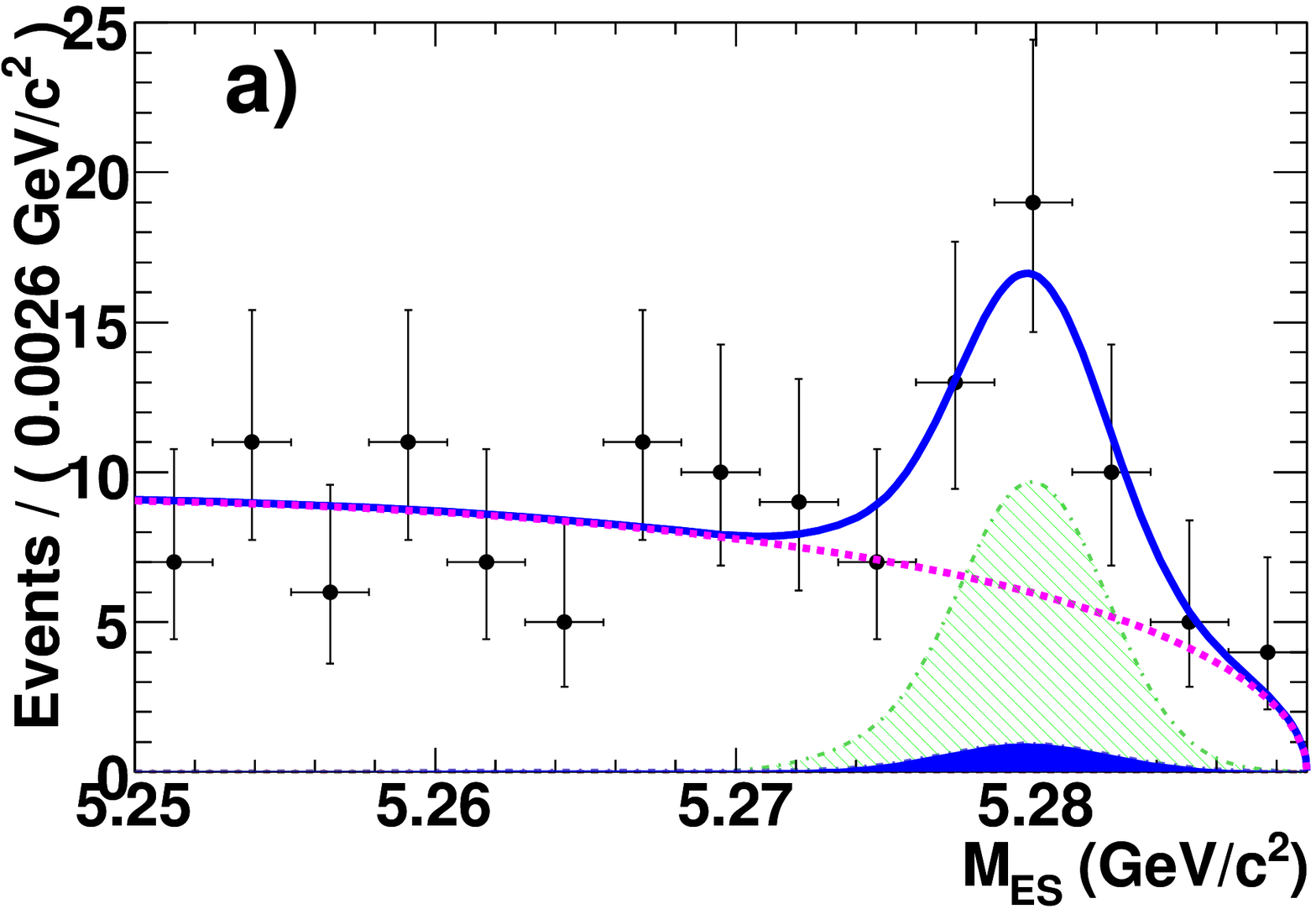}
\setlength{\epsfxsize}{0.5\linewidth}\leavevmode\epsfbox{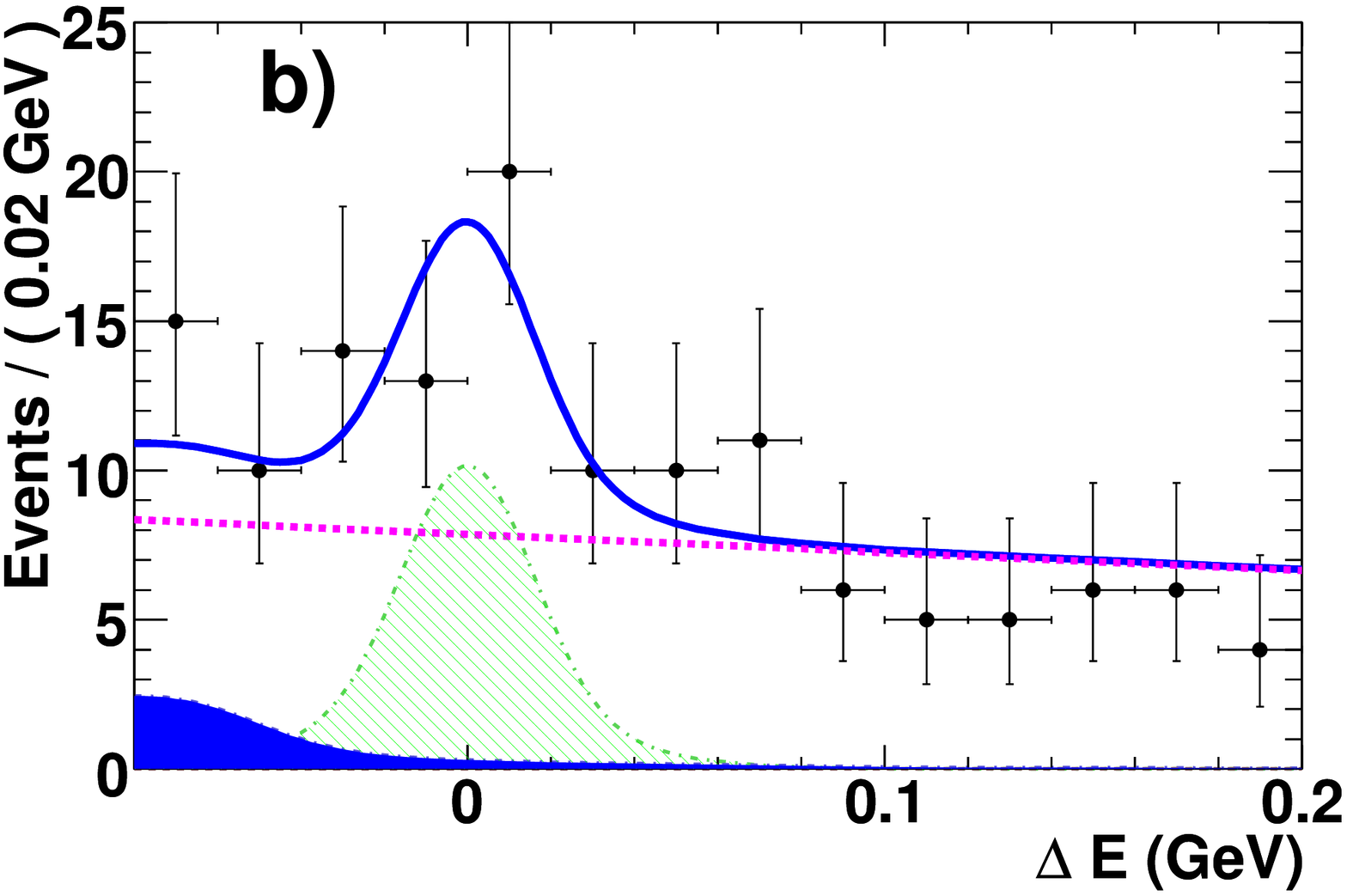}
}
\centerline{
\setlength{\epsfxsize}{0.5\linewidth}\leavevmode\epsfbox{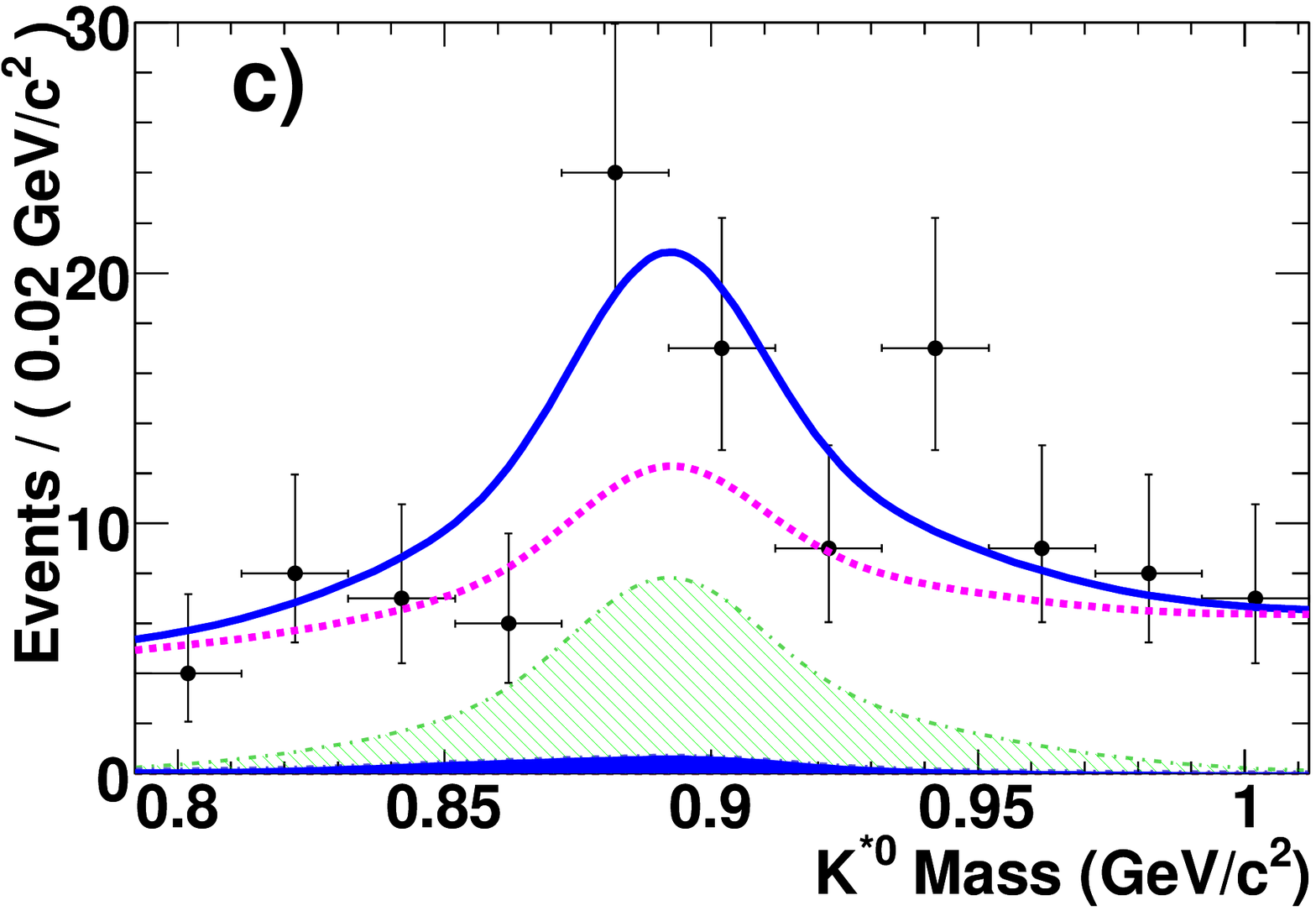}
\setlength{\epsfxsize}{0.5\linewidth}\leavevmode\epsfbox{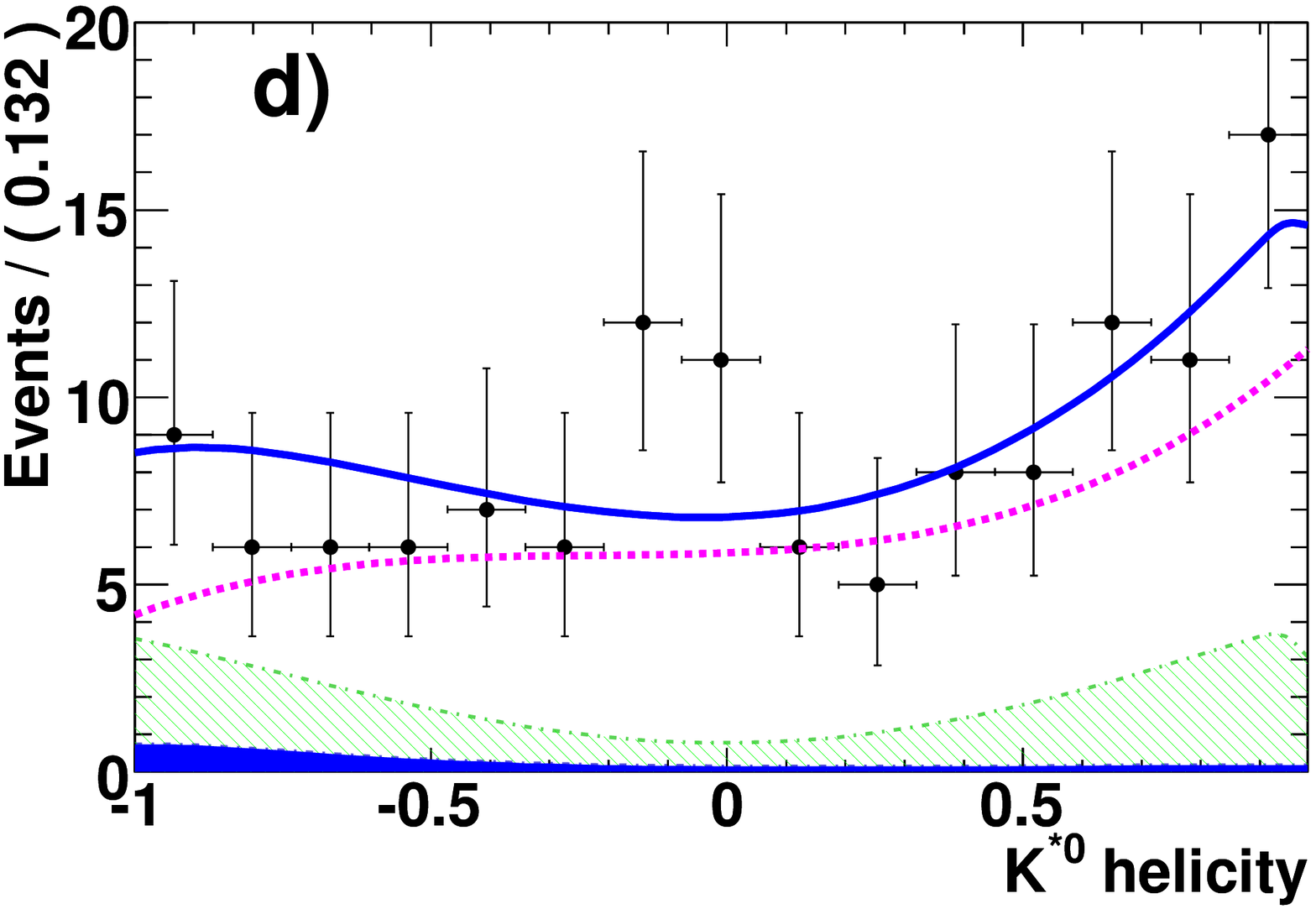}
}
\vspace{-0.3cm}
\caption{\label{fig:projections} (color online) Projections of the
multidimensional fit onto (a) \mes; (b) \DeltaE; (c) \Kstarz\ mass;
and (d) cosine of \Kstarz\ helicity angle for \btoKstarzKstarzb\
events selected with a requirement on the signal-to-total likelihood
probability ratio, optimized for each variable, with the plotted
variable excluded.  The points with error bars show the data; the
solid line shows signal-plus-background; the
dashed line is the continuum background; the hatched region is the
signal; and the shaded region is the \Bback.}
\end{figure}

The total event sample consists of 7363 and 1390 events for
\btoKstarzKstarzb\ and \btoKstarzKstarz, respectively.  The results of
the ML fits are summarized in Table~\ref{tab:results}.  The \Bback\
yield agrees with the MC prediction within the statistical errors.
The significance $S$ of the signal is defined as $S=2\Delta\ln {\cal
L}$, where $\Delta\ln {\cal L}$ is the change in likelihood from the
maximum value when the number of signal events is set to zero,
corrected for the systematic error defined below.  The robustness of
the significance estimate is cross-checked through fitting a series of
toy MC ensembles generated from the fitted parameters.  The
significance of the \btoKstarzKstarzb\ branching fraction is $\kzbsig
\sigma$, including statistical and systematic uncertainties.  For
\btoKstarzKstarz, we compute the 90\% C.L.  upper limit as the
branching fraction below which lies 90\% of the total likelihood
integral, taking into account the systematic uncertainty.
Figure~\ref{fig:projections} shows the projections of the fits onto
\mes, \DeltaE, \Kstarz\ mass and cosine of the \Kstarz\ helicity angle
for \btoKstarzKstarzb.

\begin{table*}[htb]
\caption{Summary of results: signal yield $n_{\rm sig}$, the \Bback\
yield $n_{\BB}$, signal reconstruction efficiency $\varepsilon$
(taking into account that ${\cal B}(\Kstarz\rightarrow K^+\pi^-)
=2/3$), significance $S$ (systematic uncertainties included),
branching fraction ${\cal B}$, 90\% C.L. upper limit for
\btoKstarzKstarz\ branching fraction, and the longitudinal
polarization \fL. The first error given is statistical and the second
is systematic.}
\begin{center}
\begin{tabular}{lccc}
\hline \hline 
\raisebox{-1ex}{Channel}  &
\raisebox{-1ex}{\KstarzKstarzb} & \hspace{0.5cm}
 & \raisebox{-1ex}{\KstarzKstarz} \\[3pt] \hline
$n_{\rm sig}$          & \kzbnsig & &  \kznsig \\
$n_{\BB}$          & $19 \pm 12$ & &  $68\pm29$\\
$\varepsilon$ (\%)  &  6.8 & &  6.4\\ 
$S\  (\sigma)$       &  \kzbsig  & &  \kzsig \\
${\cal B}(10^{-6})$ &  \kzbbfcor & &  \kzbfcor \\
UL ${\cal B}(10^{-6})$ &  - & &  \kzupcor\\
\fL\              &  \kzbfl & &  \kzfl \smallskip \\ 
\hline
\hline
\end{tabular}
\label{tab:results}
\end{center}
\end{table*}

Systematic uncertainties in the branching fractions are dominated by
our knowledge of the PDF modeling. Varying the
PDF parameters by their errors results in changes in the yields of
6.5\% and 19.0\% for \btoKstarzKstarzb\ and \btoKstarzKstarz,
respectively. The largest contribution comes from the width
of the \Kstarz.

\begin{figure}[htb!]
\begin{center}
  \includegraphics[width=0.8\linewidth]{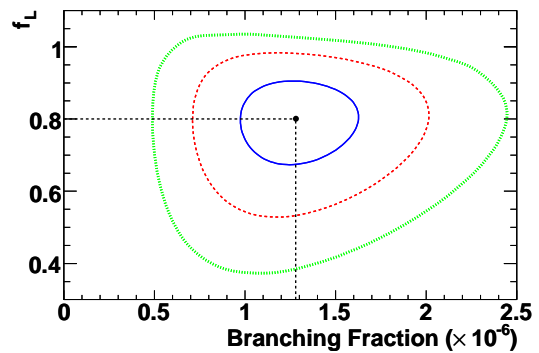} 
\end{center}
\vspace{-0.3cm}
\caption{Distribution of $-2\ln {\cal L} ({\cal B},\fL)$ for 
\btoKstarzKstarzb\ decays. The solid dot shows
 central value and the curves give contours in steps of one sigma 
($\Delta \sqrt{-2\ln {\cal L} ({\cal B},\fL)}=1$).}
\label{fig:contour}
\end{figure}

The reconstruction efficiency depends on the decay polarization.  We
calculate the efficiency using the measured polarization and assign a
systematic error from the uncertainty on \fL\ of 3.4\% and 27.0\% for
\btoKstarzKstarzb\ and \btoKstarzKstarz,
respectively. Figure~\ref{fig:contour} shows the behavior of $-2\ln
{\cal L}({\cal B}, \fL)$ for the \btoKstarzKstarzb\ mode.

The uncertainties in PDF modeling and \fL\ are additive in nature and
affect the significance of the branching fraction results.
Multiplicative uncertainties include reconstruction efficiency
uncertainties from tracking (3.2\%) and particle identification
(4.4\%), track multiplicity (1\%), MC signal efficiency statistics
(0.6\%), and the number of \BB\ pairs (1.1\%). Variation of the
expected yield from $\Bz\to\Kstarz\KstarzbII$ events has a negligible
effect on the signal.

The systematic uncertainty in \fL\ is dominated by the PDF shape
variations, which contribute 7\% for \btoKstarzKstarzb\ and 20\%
for \btoKstarzKstarz. Other errors identified
above for the branching fraction have a very small effect on
\fL\ and contribute in total 0.7\%. The total
systematic error is summarized in Table~\ref{tab:results}.


In summary, we have measured the branching fraction ${\cal B}
(\btoKstarzKstarzb) = [\kzbbfcorsys] \times 10^{-6}$ with a
significance of $\kzbsig \sigma$. We find the fraction of longitudinal
polarization \fL\ = \kzbflsys. Both results are in agreement with the
upper range of theoretical predictions. The 90\% C.L. upper limit on
the branching fraction ${\cal B} (\btoKstarzKstarz) < \kzupcor\times
10^{-6}$ is two orders of magnitude more stringent than previous
measurements.

We are grateful for the excellent luminosity and machine conditions
provided by our \pep2\ colleagues, 
and for the substantial dedicated effort from
the computing organizations that support \babar.
The collaborating institutions wish to thank 
SLAC for its support and kind hospitality. 
This work is supported by
DOE
and NSF (USA),
NSERC (Canada),
CEA and
CNRS-IN2P3
(France),
BMBF and DFG
(Germany),
INFN (Italy),
FOM (The Netherlands),
NFR (Norway),
MIST (Russia),
MEC (Spain), and
STFC (United Kingdom). 
Individuals have received support from the
Marie Curie EIF (European Union) and
the A.~P.~Sloan Foundation.


\end{document}